\documentclass[aps,prl,twocolumn,showpacs,amsmath,amssymb]{revtex4}
\usepackage{epsf}
\usepackage{graphicx}
\usepackage{bm}

\begin{document}

\title{Coulomb blockade of anyons}

\author{Dmitri V. Averin and James A. Nesteroff}

\affiliation{Department of Physics and Astronomy, Stony Brook
University, SUNY, Stony Brook, NY 11794-3800 }

\date{\today}

\begin{abstract}
Coulomb interaction turns anyonic quasiparticles of a primary
quantum Hall liquid with filling factor $\nu =1/(2m+1)$ into
hard-core anyons. We have developed a model of coherent transport of
such quasiparticles in systems of multiple antidots by extending the
Wigner-Jordan description of 1D abelian anyons to tunneling
problems. We show that the anyonic exchange statistics manifests
itself in tunneling conductance even in the absence of quasiparticle
exchanges. In particular, it can be seen as a non-vanishing resonant
peak associated with quasiparticle tunneling through a line of three
antidots.

\end{abstract}


\maketitle

Quasiparticles of two-dimensional (2D) electron liquids in the
regime of the Fractional Quantum Hall effect (FQHE) have unusual
properties of fractional charge \cite{b1} and fractional exchange
statistics \cite{b2,b3}. The fractional quasiparticle charge was
observed in experiments on antidot tunneling \cite{ant1} and
shot-noise measurements \cite{b4,b5}. The situation with fractional
statistics is so far less certain even in the case of the abelian
statistics, which is the subject of this work. Although the recent
experiments \cite{int1} demonstrating unusual flux periodicity of
conductance of a quasiparticle interferometer can be interpreted as
a manifestation of the fractional statistics \cite{int2,int3}, this
interpretation is not universally accepted \cite{int4,int5}. There
is a number of theoretical proposals (see, e.g., \cite{kf2,kf4})
suggesting tunnel structures where the statistics manifests itself
through noise properties. Partly due to complexity of noise
measurements, such experiments have not been performed successfully
up to now. In this work, we show that coherent quasiparticle
dynamics in multi-antidot structures should provide clear signatures
of the exchange statistics in dc transport. Most notably, in
tunneling through a line of three antidots, fractional statistics
leads to a non-vanishing peak of the tunnel conductance which would
vanish for integer statistics.

\begin{figure}
\setlength{\unitlength}{1.0in}
\begin{picture}(3.2,1.3)
\put(.45,-.1){\epsfxsize=2.3in \epsfbox{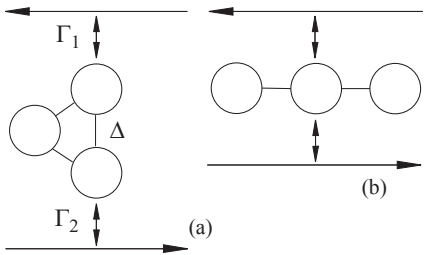}}
\end{picture}
\caption{Tunneling of anyonic quasiparticles between opposite edges
of an FQHE liquid through quasi-1D triple-antidot systems: (a) loop,
(b) open interval. Quasiparticles tunnel between the edges and the
antidots with rates $\Gamma_{1,2}$. The antidots are coupled
coherently by tunnel amplitudes $\Delta$. } \label{f1}
\end{figure}

These effects rely on the ability of quantum antidots to localize
individual quasiparticles of the QH liquids \cite{ant1,ant2,ant3}.
The resulting transport phenomena in antidots are very similar to
those associated with the Coulomb blockade \cite{al} in tunneling
of individual electrons in dots. For instance, similarly to a
quantum dot \cite{dot}, the linear conductance of one antidot
shows periodic oscillations with each period corresponding to the
addition of one quasiparticle \cite{ant1,ant2,ant3,ant4,ant5}.
Recently, we have developed a theory of such Coulomb-blockade-type
tunneling for a double-antidot system \cite{we1}, where
quasiparticle exchange statistics does not affect the transport.
The goal of this work is to extend this theory to antidot
structures where the statistics does affect the conductance. The
two simplest structures with this property consist of three
antidots and have quasi-1D geometries with either periodic or open
boundary conditions (Fig.~\ref{f1}). A technical issue that needed
to be resolved to calculate the tunnel conductance is that the
anyonic field operators defined through the Wigner-Jordan
transformation \cite{an1,an2,an3,an4}, are not fully sufficient in
the situations of tunneling. As we show below, to obtain correct
matrix elements for anyon tunneling, one needs to keep track of
the appropriate boundary conditions of the wavefunctions which are
not accounted for in the field operators.

Specifically, we consider the antidots coupled by tunneling among
themselves and to two opposite edges of the quantum Hall liquid
(Fig.~\ref{f1}). The edges play the role of the quasiparticle
reservoirs with the transport voltage $V$ applied between them. We
assume that the antidot-edge coupling is weak and can be treated
as a perturbation. Quasiparticle transport through the antidots is
governed then by the kinetic equation similar to that for
Coulomb-blockade transport through quantum dots with a discrete
energy spectrum \cite{b8}. Coherent quasiparticle dynamics
requires that the relaxation rate $\Gamma_d$ created by direct
Coulomb antidot-edge coupling is weak. This condition should be
satisfied if the edge-state confinement is sufficiently strong
\cite{we1}. The requirement on the confinement is less stringent
in the case of the antidot line (Fig.~1b), in which antidot
quasiparticles move along the edge, suppressing the antidot-edge
coupling at low frequencies. We also assume that all quasiparticle
energies on the antidots, tunnel amplitudes $\Delta$, temperature
$T$, Coulomb interaction energies $U$ between quasiparticles on
different antidots, are much smaller than the energy gap
$\Delta^*$ for excitations on each antidot. This condition ensures
that the state of each antidot is characterized completely by the
occupation number $n$ of its relevant quantized energy level. In
any given range of the backgate voltage or magnetic field (which
produces the overall shift of the antidot energies - see, e.g.,
\cite{ant1,ant2,ant3}), there can be at most one quasiparticle on
each antidot, $n=0,1$. This ``hard-core'' property of the
quasiparticles means that they behave as fermions in terms of
their occupation factors, despite the anyonic exchange statistics.
All these assumptions can be summarized as: $ \Gamma_{d}, \Gamma_j
\ll \Delta, U, T \ll \Delta^*$.

Under these conditions, the antidot tunneling is dominated by the
antidot energies. The quasi-1D geometry of the antidot systems we
consider makes it possible to introduce the quasiparticle
``coordinate'' $x$ numbering successive antidots; e.g., $x=-1,0,1$
for systems in Fig.~\ref{f1}. The quasiparticle Hamiltonian can be
the written as
\begin{equation}
H=\sum_x [ \epsilon_x n_x -(\Delta_x \xi_{x+1}^{\dagger}\xi_x +
h.c.)] + \sum_{x< y} U_{x,y} n_x n_y \, ,  \label{e2}
\end{equation}
where $\epsilon_x$ are the energies of the relevant localized states
on the antidots (taken relative to the common chemical potential of
the edges at $V=0$), $\Delta_x$ is the tunnel coupling between them,
$U_{x,y}$ is the quasiparticle Coulomb repulsion, and $n_x \equiv
\xi_x^{\dagger}\xi_x$. The quasiparticle operators $\xi_x^{\dagger},
\xi_x$ in (\ref{e2}) can be viewed as the Klein factors left in the
standard operators for the edge-state quasiparticles when all the
edge magneto-plasmon modes are suppressed by the gap $\Delta^*$.
Characteristics of such Klein factors depend on the geometry of a
specific tunneling problem; non-trivial examples can be found in
\cite{kf2,kf4,kf1,kf3}. In the Hamiltonian (\ref{e2}), $\xi_x$
describe the hard-core anyons with exchange statistics $\pi \nu$.
Wigner-Jordan transformation expresses them through the Fermi
operators $c_x$ \cite{an1}:
\begin{equation}
\xi_x = e^{i \pi (\nu-1) \sum_{z<x} n_z } c_x \, , \; \xi_y \xi_x=
\xi_x \xi_y e^{i \pi \nu sgn (x-y)} , \label{e3}
\end{equation}
with similar relations for $\xi^{\dagger}$.

Anyonic statistics creates an effective interaction between the
quasiparticles which can be understood as the Aharonov-Bohm (AB)
interaction between a flux tube ``attached'' to one of the particles
and the charge carried by another. In general, this interaction can
be masked by the direct Coulomb interaction $U_{x,y}$. In the {\em
antidot loop} (Fig.~\ref{f1}a), however,  $U_{x,y}$ is constant,
$U_{x,y}=U$, and the interaction term in (\ref{e2}) reduces to
$Un(n-1)/2$, with $n=\sum_x n_x$ -- the total number of the
quasiparticles on the antidots. In this case, the Coulomb
interaction contributes to the energy separation between the group
of states with different $n$, but does not affect the level
structure for given $n$. The hard-core property of quasiparticles
limits $n$ to the interval $[0,3]$. For $n=0$ and $n=3$, the system
has the ``empty'' and ``completely filled'' state with respective
energies $E_0=0\, , E_3=\sum_x \epsilon_x +3U$. The spectrum
$E_{1k}$ of the three $n=1$ states $|1k\rangle=\sum_x \phi_k(x)
\xi_x^{\dagger} |0\rangle $, is obtained as usual from (\ref{e2}).
In the uniform case $\epsilon_x= \epsilon$, $\Delta_x= \Delta $,
with an external AB phase $\varphi$, one has $\phi_k
(x)=e^{ikx}/L^{1/2}$ and
\begin{equation}
E_{1k}= \epsilon -\Delta \cos k\, , \; k=(2\pi m +\varphi)/L \, ,
\label{e6} \end{equation} where $m=0,1,2$, and the loop length is
$L=3$.

\begin{figure}[h]
\setlength{\unitlength}{1.0in}
\begin{picture}(3.3,0.4)
\put(.0,-.15){\epsfxsize=3.3in \epsfbox{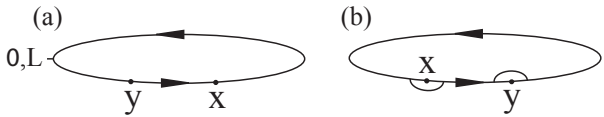}}
\end{picture}
\caption{Exchanges of hard-core anyons on a 1D loop: (a) real
exchanges by transfer along the loop embedded in a
2D system; (b) formal exchanges describing the assumed boundary
conditions \protect (\ref{e9}) of the wavefunction. } \label{f2}
\end{figure}

Anyonic statistics can be seen in the $n=2$ states, $|2l
\rangle=(1/\sqrt{2})\sum_{xy} \psi_l(x,y) \xi_y^{\dagger}
\xi_x^{\dagger} |0\rangle$. The fermion-anyon relation (\ref{e3})
suggests that the stationary two-quasiparticle wavefunctions
should coincide up to the exchange phase with that for free
fermions:
\begin{equation}
\psi_l(x,y) =\frac{e^{i \pi (1- \nu) sgn (x-y)/2} }{\sqrt{2}} \det
\left( \begin{array}{cc} \phi_q (x) & \phi_q (y) \\ \phi_p (x) &
\phi_p (y) \end{array} \right) . \label{e8}  \end{equation} Here
$\phi$s are the single-particle eigenstates of the Hamiltonian
(\ref{e2}). (The states (\ref{e8}) are numbered with the index $l$
of the third ``unoccupied'' eigenstate of (\ref{e2}) complementary
to the two occupied ones $q,p$.) The boundary conditions for the
$\phi$s are affected by the exchange phase in Eq.~(\ref{e8}). To
find them, we temporarily assume for clarity that coordinates
$x,y$ are continuous and lie in the interval $[0,L]$. Subsequent
discretization does not change anything substantive in this
discussion. The 1D hard-core particles are impenetrable and can be
exchanged only by moving one of them, say $x$, around the loop
from $x=y+0$ to $x=y-0$ (Fig.~\ref{f2}a). Since the loop is
imbedded in the underlying 2D system, such an exchange means that
the wavefunction acquires the phase factor $e^{i\pi \nu}$, in
which the sign of $\nu$ is fixed by the properties of the 2D
system, e.g. the direction of magnetic field in the case of FQHE
liquid.  Next, if the second particle is moved similarly, from
$y=x+0$ to $y=x-0$, the wavefunction changes in the same way, for
a total factor $e^{i 2 \pi \nu}$. Equation (\ref{e8}) shows that
only one of these changes can agree with the 1D form of the
exchange phase. As a result, the wavefunction (\ref{e8}) satisfies
different boundary conditions in $x$ and $y$:
\begin{equation}
\psi_l(L,y) =\psi_l(0,y) e^{i \varphi}, \;
\psi_l(x,L)=\psi_l(x,0)e^{i (\varphi +2 \pi \nu) }  . \label{e9}
\end{equation} Conditions (\ref{e9}) on the wavefunction (\ref{e8})
mean that the single-particle functions $\phi$ in (\ref{e8})
satisfy the boundary condition that correspond to the effective AB
phase $\varphi'= \varphi+ \pi -\pi \nu$, i.e. the addition of an
extra quasiparticle to the loop changed the AB phase by $\pi -\pi
\nu $, where $- \pi \nu$ comes from the exchange statistics and
$\pi$ from the hard-core condition. This gives the energies of the
two-quasiparticle states (\ref{e8}) as $U+E_{1q}+E_{1p}$, where,
if the loop is uniform, the single-particle energies are given by
Eq.~(\ref{e6}) with $\varphi \rightarrow \varphi'$. In this case,
$\sum_k E_{1k}=0$, and the energies $E_{2l}$ of the
two-quasiparticle states can be written as:
\begin{equation}
E_{2l}= 2\epsilon +U-\Delta \cos l\, , \; l=(2\pi m' +\varphi- \pi
\nu)/3 \, , \label{e10} \end{equation} where $m'=0,1,2$.

One of the consequences of this discussion is that the sign of
$\nu$ in the 1D exchange phases of Eqs.~(\ref{e3}) and (\ref{e8})
can be chosen arbitrarily for a given fixed sign of the 2D exchange
phase. Reversing this sign only exchanges the character of the
boundary conditions (\ref{e9}) between $x$ and $y$. This fact has
simple interpretation. Although the 1D hard-core anyons can not be
exchanged directly, formally, coordinates $x$ and $y$ in
Eq.~(\ref{e8}) are independent and one needs to define how they
move past each other at the point $x=y$. Depending on whether the
$x$-particle moves around $y$ from below or (as in Fig.~\ref{f2}b)
from above, its trajectory does or does not encircle the flux
carried by the $y$ particle, and the boundary condition for $x$ is
or is not affected by the statistical phase. The choice made for $x$
immediately implies the opposite choice for $y$ (Fig.~\ref{f2}b),
accounting for different boundary conditions (\ref{e9}). This
interpretation shows that in calculation of any matrix elements, the
participating wavefunctions should be taken to have the same
boundary conditions. While this requirement is natural for processes
with the same number of anyons, it is less evident for tunneling
that changes the number of anyons in the system. Indeed, the most
basic, tunnel-Hamiltonian, description of tunneling into the point
$z$ of the system leads to the states
\begin{equation}
\xi_z^{\dagger} |1k \rangle = (1/\sqrt{2})\sum_{xy} \psi_k (x,y)
\xi_y^{\dagger} \xi_x^{\dagger} |0\rangle \, , \label{e11}
\end{equation}

\vspace*{-2ex}

\[ \psi_k (x,y) = [ \phi_k(x) \delta_{y,z} - e^{i
\pi (1- \nu) sgn (x-y)} \delta_{x,z} \phi_k(y)] / \sqrt{2} \, . \]
One can see that Eq.~(\ref{e11}) automatically implies specific
choice of the boundary conditions which corresponds to the tunneling
anyon not being encircled by anyons already in the system. This
means that in the calculation of the tunnel matrix elements with the
states (\ref{e8}), one should always pair the coordinate of the
tunneling anyon with the discontinuous one in (\ref{e9}). Then, the
tunnel matrix elements are obtained as
\begin{equation}
\langle 2l| \xi_z^{\dagger} |1k \rangle= \sqrt{2} \sum_x
\psi^*_l(x,z)\phi_k(x)\, . \label{e12}
\end{equation}
For instance, in the case of uniform loop with states (\ref{e6}) and
(\ref{e10}), we get up to an irrelevant phase factor
\begin{equation}
\langle 2l| \xi_z^{\dagger} |1k \rangle= (2 /3) \cos [(k-l)/2]\, .
\label{e13}
\end{equation}
Specific anyonic interaction between quasiparticles can be seen in
the fact that the matrix elements (\ref{e13}) do not vanish for
any pair of indices $k,l$. In the fermionic case $\nu=1$, one of
the elements (\ref{e13}) always vanishes for any given $k$, since
the two-particle state after tunneling necessarily has one
particle in the original single-particle state. By contrast, the
tunneling anyon can shift existing particle out of its state.

The matrix elements involving empty or fully occupied states
coincide with those for fermions. Taken together with
Eqs.~(\ref{e12}) and (\ref{e13}) for transitions between the
partially filled states, they determine the rates $\Gamma_j (E) =
\gamma_j f_{\nu} (E) |\langle \xi_z^{\dagger} \rangle |^2$ of
tunneling between the $j$th edge and the antidots, where
$\gamma_j$ is the overall magnitude of the tunneling rate, and
\[f_{\nu} (E) =(2\pi T/\omega_c)^{\nu -1} | \Gamma (\nu /2 + i
E/2\pi T )|^2 e^{-E/2T}/2\pi \Gamma(\nu) \] is its energy dependence
associated with the Luttinger-liquid correlations in the edges
\cite{b6}. Here $\Gamma(z)$ is the gamma-function and $\omega_c$ is
the cut-off energy of the edge excitations.  The rates $\Gamma_j(E)$
can be used in the standard kinetic equation to calculate the
conductance of the antidot system \cite{we1}. Anyonic statistics of
quasiparticles affects the position and amplitude of the conductance
peaks through the shift of the energy levels by quasiparticle
tunneling (described, e.g., by Eq.~(\ref{e10})) and through the
kinetic effects caused by the anyonic features in the matrix
elements (\ref{e12}). In the case of the antidot loop
(Fig.~\ref{f1}a), however, effects of statistics are masked by the
external AB flux $\varphi$ through the loop. Since the area of
practical antidots is much larger than the internal area of the
loop, $\varphi$ is essentially random and can not be controlled by
external magnetic field on the relevant scale of one period of
conductance oscillations. Below, we present the results for
conductance for the similar case of a {\em line of antidots}
(Fig.~\ref{f1}b), the conductance of which is insensitive to the AB
flux, and shows effects of fractional statistics in the tunneling
matrix elements.

\begin{figure}
\setlength{\unitlength}{1.0in}
\begin{picture}(2.5,1.5)
\put(-.3,-.15){\epsfxsize2.5in \epsfbox{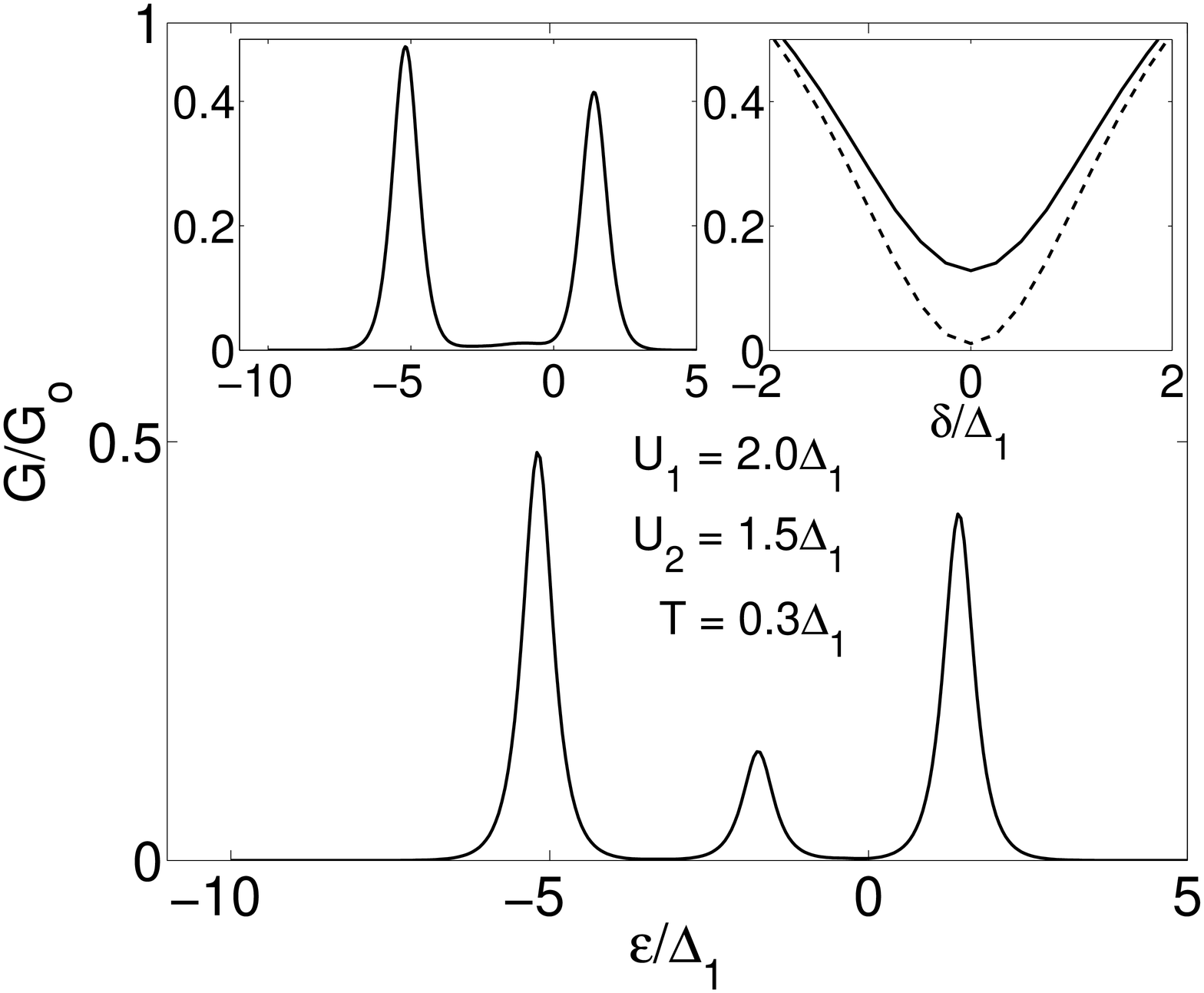}}
\end{picture}
\caption{Linear conductance $G$ of the antidot line in a $\nu=1/3$
FQHE liquid (Fig.~\ref{f1}b) as a function of the common antidot
energy $\epsilon$ relative to the edges. In contrast to electrons
($\nu=1$, left inset), tunneling of quasiparticles with fractional
exchange statistics produces non-vanishing conductance peak
associated with transition between the ground states of one and two
quasiparticles. The maximum of this peak is shown in the right inset
($\nu = 1/3$ -- solid, $\nu =1$ -- dashed line) as a function of the
bias $\delta$. The curves are plotted for $\Delta_1 = \Delta_2$,
$\lambda=0$, $\gamma_1=\gamma_2$; conductance is normalized to $G_0
= (e\nu)^2\Gamma_1(0)/\Delta_1$. } \label{f3}
\end{figure}

As before, the quasiparticle Hamiltonian is given by Eq.~(\ref{e2}).
In this geometry, the interaction energy $U_1 \equiv U_{1,0} =
U_{0,-1}$ between the nearest-neighbor antidots is in general
different from the interaction $U_2 \equiv U_{1,-1}$ between the
quasiparticles at the ends. The localization energies on the
antidots can be written as $\epsilon_{j} = \epsilon + x \delta +2
\lambda |x|$. We consider first the unbiased line, $\delta=0$. At
low temperatures, $T \ll \Delta, U$, only the ground states of $n$
quasiparticles with energies $E_n$ participate in transport:
$E_0=0$, $E_1 = \epsilon + \lambda - \omega $, $E_2 = 2 \epsilon + 3
\lambda - \bar{\omega } +(U_a+U_{b})/2 $, and $E_3 = 3\epsilon +
2U_{a}+U_{b}+4\lambda$, where $\omega=(\Delta_1^2 + \Delta_2^2
+\lambda^2)^{1/2}$ and $\bar{\omega }$ is given by the same
expression with $\lambda$ replaced by $\bar{\lambda} =\lambda -
(U_1-U_2)/2$. In this regime, the linear conductance $G$ consists of
three peaks, with each peak associated with addition of one more
quasiparticle to the antidots,
\begin{equation}
G = \frac{(e\nu)^2}{T}\frac{\gamma_1 \gamma_2}{\gamma_1+\gamma_2}
\frac{ a_n f_{\nu}(E_{n+1}-E_n)}{1+\exp [-(E_{n+1}-E_n)/T]}\, ,
\label{e14} \end{equation} where $a_n \equiv |\langle n+1
|\xi_0^{\dagger}|n \rangle |^2$. The amplitudes $a_0$, $a_2$ are
effectively single-particle, and thus, independent of the exchange
statistics: $a_0=(\omega + \lambda)/ 2\omega$, and
$a_2=(\bar{\omega }- \bar{\lambda})/2\bar{\omega }$. By contrast,
the amplitude $a_1$ of the transition from one to two
quasiparticles is multi-particle, and is found from
Eqs.~(\ref{e8}) and (\ref{e12}) to be strongly
statistics-dependent,
\begin{equation}
a_1= \frac{\Delta_1^2 \Delta_2^2 }{(\omega + \lambda) \omega
(\bar{\omega }- \bar{\lambda}) \bar{\omega }} \cos^2(\pi \nu/2)\, .
\label{e15} \end{equation} In particular, $a_1$ vanishes for
electron tunneling ($\nu=1$), but is non-vanishing in the case of
fractional statistics, e.g., for $\nu=1/3$, when $\cos^2(\pi
\nu/2)=3/4$. This is illustrated in Fig.~\ref{f3} which shows the
conductance $G$ obtained by direct solution of the full kinetic
equation for tunneling through the antidots. Qualitatively, the
vanishing amplitude $a_1$ for electrons can be understood as a
result of destructive interference between the two terms in the
two-particle wavefunction which correspond to different ordering of
the added/existing electron on the antidot line. The opposite signs
of these two terms lead to vanishing overlap with the
single-particle state in the tunnel matrix element. Fractional
statistics of quasiparticles makes this destructive interference
incomplete. Finite bias $\delta \neq 0$ along the line suppresses
this interference making the effect of the statistics smaller. One
can still distinguish the fractional statistics by looking at the
dependence of the amplitude of the middle peak of conductance on the
bias $\delta$ shown in the right inset in Fig.~\ref{f3}.

In conclusion, we have developed a model of coherent transport of
anyonic quasiparticles in systems of multiple antidots. In antidot
loops, addition of individual quasiparticles shifts the
quasiparticle energy spectrum by adding statistical flux to the
loop. In the case without loops, energy levels are insensitive to
quasiparticle statistics, but the statistics still manifests
itself in the quasiparticle tunneling rates and hence dc tunnel
conductance of the antidot system.

The authors would like to thank  F.E. Camino, V.J. Goldman, J.K. Jain,
V.E. Korepin, Yu.V. Nazarov, O.I. Patu, V.V. Ponomarenko, and J.J.M.
Verbaarschot for discussions. This work was supported in part by NSF
grant \# DMR-0325551 and by ARO grant \# DAAD19-03-1-0126.

\end{document}